# Evidence for Magnetic Ordering Associated with Metal-Insulator Transition in SmRu$_4$P$_{12}$ Studied by Muon Spin Relaxation

Kenichi HACHITANI,* Hideto FUKAZAWA,† and Yoh KOHORI†

*Graduate School of Science and Technology, Chiba University,*
*1-33, Yayoi-cho, Inage-ku, Chiba, 263-8522, Japan*

Isao WATANABE

*Advanced Meson Science Laboratory, RIKEN (The Institute of Physical and Chemical Research),*
*2-1, Hirosawa, Wako, Saitama, 351-0198, Japan*

Chihiro SEKINE and Ichimin SHIROTANI

*Department of Electrical and Electronic Engineering, Muroran Institute of Technology,*
*27-1, Mizumoto-cho, Muroran, Hokkaido, 050-8585, Japan*

(Dated: October 16, 2006)

Muon spin relaxation measurements on the filled skutterudite compound SmRu$_4$P$_{12}$ have been carried out in zero- and longitudinal-fields. Temperature dependence of both initial asymmetry and muon-spin depolarization rate show the appearance of a magnetically ordered state below the metal-insulator transition temperature $T_{\rm MI}$. The present study indicates that the ordering below $T_{\rm MI}$ is not a non-magnetic antiferro-quadrupolar ordering but a magnetic one, which supports a scenario that a magnetic octupolar ordering occurs below $T_{\rm MI}$ in SmRu$_4$P$_{12}$.

## I. INTRODUCTION

Filled skutterudite compounds with the general formula $RT_4X_{12}$ ($R$: rare earth, actinoid; $T$: Fe, Ru, Os; $X$: P, As, Sb) crystallize in a body-centered cubic structure of the space group $Im\bar{3}$ ($T_h^5$, No. 204) [1]. A lot of remarkable phenomena have been observed in these systems [2–10]. Among them, PrRu$_4$P$_{12}$ and SmRu$_4$P$_{12}$ exhibit metal-insulator (M-I) transitions at $T_{\rm MI}$ of 62 K and 16.5 K, respectively [5, 11]. They have attracted much attention because of their different features of each M-I transition. PrRu$_4$P$_{12}$ shows no sign of a magnetic ordering below $T_{\rm MI}$; on the contrary, a magnetization anomaly was observed in SmRu$_4$P$_{12}$ at $T_{\rm MI}$ [12].

The nesting of the Fermi surface may be a common feature of the $R$Ru$_4$P$_{12}$ system. Indeed, band structure calculations have shown the existence of the approximately cubed-shaped holelike Fermi surface with the nesting wave vector $\boldsymbol{q} = (1, 0, 0)$ [13, 14]. The band has a roughly flat dispersion at the Fermi level, which indicates that the slight displacement causes a substantial difference in the Fermi surface topologies. The interplay of the Fermi surface instability and the 4f orbital degree of freedom which is coupled to the local lattice distortion is an unique character in this system. The lattice distortion lifts an angular momentum degeneracy within the cubic structure, and sometimes induces a multipolar ordering [14].

In PrRu$_4$P$_{12}$, lattice distortions open a gap everywhere on the Fermi surface [13, 15]. However, the structural phase transition at $T_{\rm MI}$ does not change the local symmetry of the Pr 4f state, which makes the system remaining in a paramagnetic state. In SmRu$_4$P$_{12}$, magnetic susceptibility measurements show that Sm ions are in a trivalent state with the total angular momentum of $J = 5/2$ [11]. In the body-centered cubic structure with the point group of $T_h$ symmetry, the crystalline electric field splits a 6-fold degenerate multiplet into the doublet $\Gamma_5$ and the quartet $\Gamma_{67}$. Specific heat measurements show that the magnetic entropy at $T_{\rm MI}$ is close to $R\ln 4$ [16]. This result means that the ground state is $\Gamma_{67}$ which corresponds to $\Gamma_8$ in the ordinary $O_h$ symmetry having both a magnetic and an orbital degrees of freedom.

In SmRu$_4$P$_{12}$, an additional phase transition at $T_{\rm N}$ of 15 K was observed below $T_{\rm MI}$. The $T_{\rm MI}$ exhibits the characteristic magnetic field dependence [17]. The $T_{\rm MI}$ increases with increasing field up to 200 kOe, and then the field dependence saturates at fields up to 300 kOe. The reentrant behavior is expected at further high fields. On the contrary, $T_{\rm N}$ decreases with increasing field. Though the anomaly at $T_{\rm N}$ is vague in zero field (ZF), it becomes apparent with increasing field [12, 17]. The behavior is quite similar to that in CeB$_6$ where an antiferroquadrupolar (AFQ) ordering and a following antiferromagnetic (AFM) one occur [18]. Hence, the successive transition in SmRu$_4$P$_{12}$ has been expected to be an AFQ ordering below $T_{\rm MI}$ and an AFM one below $T_{\rm N}$ [12, 17].

---

*Also at Advanced Meson Science Laboratory, RIKEN; Electronic address: hachitani@physics.s.chiba-u.ac.jp
†Also at Department of Physics, Faculty of Science, Chiba University

On the other hand, recent ultrasonic measurements have shown that the elastic constants $C_{44}$ and $\frac{1}{2}(C_{11}-C_{12})$ exhibit a slight and a large elastic softenings above $T_{MI}$ and below $T_{MI}$ toward $T_N$, respectively [19, 20]. The softening above $T_{MI}$ is much less than that expected for the case of an AFQ ordering in which the quadrupole-quadrupole interaction plays an important role [20]. From group-theoretical considerations, an octupolar ordering ($\Gamma_{5u}$) below $T_{MI}$ and an AFM one ($\Gamma_{4u}$) below $T_N$ is considered as the most probable candidates [20]. To date, an octupolar ordering in $NpO_2$ is widely accepted [21–23]. $SmRu_4P_{12}$ would be a following candidate.

In ZF, there is a clear difference between an AFQ ordering and an octupolar one. An AFQ ordering is non-magnetic and holds the time reversal symmetry (TRS); on the contrary, an octupolar one is magnetic and breaks TRS. In the case of an AFQ ordering in an applied magnetic field, it is noted that TRS can be broken by the appearance of an internal field from an AFM component induced by the external field. In the case of an octupolar ordering, TRS is broken even in an applied field because it is essentially a magnetic ordering. As a result, TRS is broken in both cases in an applied field. It is impossible to distinguish a magnetically field induced AFQ ordering from an magnetic octupolar one under an external field. A magnetization anomaly at $T_{MI}$ in $SmRu_4P_{12}$ was observed in the applied fields of 10-90 kOe [12]. Hence, further microscopic ZF-measurements are strongly required to clarify whether the phase transition at $T_{MI}$ is "non-magnetic" or "magnetic". The neutron diffraction experiment on this system is difficult because of the large absorption of neutrons by Sm nuclei. Though a $^{31}$P-NMR measurement has already been performed [24], the information in ZF could not be obtained since this experiment can not be performed in ZF. The muon spin relaxation ($\mu$SR) measurement is the best method for the present purpose because the $\mu$SR experiment can be performed in ZF and sensitively detect only magnetic components. Therefore, ZF-$\mu$SR measurements on $SmRu_4P_{12}$ have been carried out in order to clarify the transition at $T_{MI}$. In this letter, we report the results of the $\mu$SR measurements in ZF and longitudinal fields (LF).

## II. EXPERIMENTAL

The single-phase polycrystalline $SmRu_4P_{12}$ was synthesized by using the high pressure and high temperature method [11]. The sample was crushed to powder for the measurements. The $\mu$SR experiments were performed by implanting pulsed surface positive muons at the RIKEN-RAL Muon Facility in the UK. The direction of the initial muon spin is parallel to the beam line. Forward and backward counters were located on the upstream and the downstream sides in the direction of the beam line. The asymmetry parameter $A(t)$ was defined as $A(t) = [F(t) - \alpha B(t)]/[F(t) + \alpha B(t)] - A_{BG}$, where $F(t)$ and $B(t)$ are the total muon events counted by the forward and the backward counters at time $t$, respectively. The $A_{BG}$ is the background. The $\alpha$ is the calibration factor reflecting the relative counting efficiencies of both counters. In the ZF-experiments, stray fields at the sample position were compensated within 0.05 Oe by using correction coils, which is small enough for our ZF-$\mu$SR measurement. The LF-$\mu$SR experiments were performed by applying magnetic fields up to 500 Oe along the beam line.

## III. RESULTS AND DISCUSSION

Figure 1 shows the ZF-$\mu$SR time spectra at several temperatures $T$. The $A(t)$ gradually decreases with $t$ above $T_{MI}$, where the muon-spin depolarizes slowly by transferred magnetic fields from dynamically fluctuating Sm moments and nuclear moments. The decrease of $A(t)$ becomes rapid below about 17 K. The muon-spin precession was observed at low temperatures below about 5 K, which indicates that the ground state of the Sm moments is a magnetically ordered state.

In order to clearly show the change of the spectrum shown in FIG. 1, the spectra were analyzed by a multi-components function expressed by the formula

$$A(t) = A_1 e^{-\lambda_1 t} + A_2 e^{-\lambda_2 t} + A_3 e^{-\lambda_3 t}\cos(\omega t + \theta). \quad (1)$$

The first and the second terms simply represent the components of the rapid and the slow depolarizations, respectively. The third term describes the precession component. The parameters of $A_1$, $A_2$, $A_3$ are the initial asymmetries at $t = 0$, the $\lambda_1$, $\lambda_2$, $\lambda_3$ are the muon-spin depolarization rates, and $\omega$, $\theta$ are the frequency and the phase of the precession. In the analysis of the spectra below about 5 K which includes the muon-spin precession component, the third term was taken into account. The spectra above the temperature which have no precession were analyzed without this term.

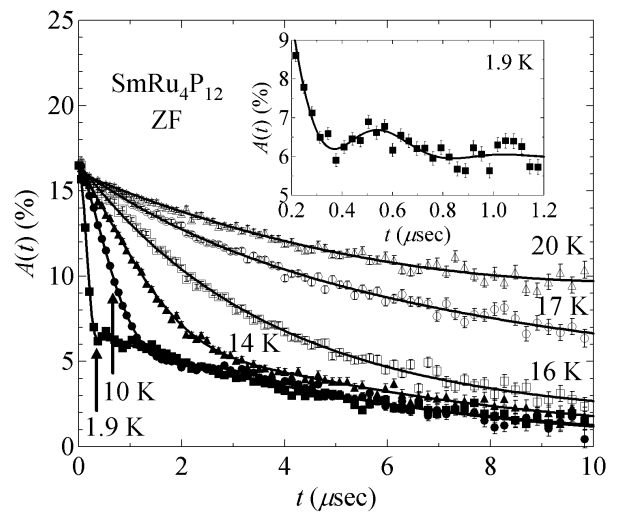

FIG. 1: ZF-$\mu$SR time spectra at several temperatures. The inset shows the muon-spin precession at 1.9 K. The solid lines show the best-fit results by the formula (1).

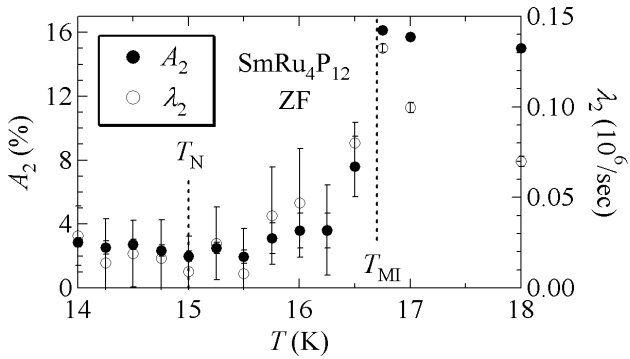

FIG. 2: Temperature dependence of $A_2$ (closed circles) and $\lambda_2$ (open circles) obtained from the best-fit results by the formula (1) shown as the solid lines in FIG. 1.

Figure 2 shows $T$ dependence of $A_2$ and $\lambda_2$. At $T_{\rm MI}$, $A_2$ rapidly decreases, and $\lambda_2$ shows a peak. The decrease of $A_2$ means the increase of $A_1$, which corresponds to the appearance of the fast relaxation behavior due to a magnetic origin. The peak in $\lambda_2$ arises from the critical slowing down behavior of the Sm moments around $T_{\rm MI}$. These results strongly support the appearance of a magnetically ordered state below $T_{\rm MI}$ [25, 26]. On the contrary, no remarkable anomaly was observed around $T_{\rm N}$.

Since the muon-spin precession was not clearly observed above about 5 K, LF-$\mu$SR measurements were performed at several temperatures in order to clarify whether the anomaly at $T_{\rm MI}$ is due to a static or a dynamical internal field. Figure 3 shows the LF-$\mu$SR time spectra at 1.9 K in several longitudinal fields $H_{\rm LF}$. The tail of the spectrum longer than about 1 $\mu$sec is lifting up with increasing $H_{\rm LF}$. The spectral overlap exists below about 0.3 $\mu$sec. The behavior is typical in the presence of a static internal field at the muon site [27]. Long-term depolarizations were observed in each field due to the contribution from the fluctuations of the Sm moments at 1.9 K.

Figure 4 shows $H_{\rm LF}$ dependence of $A_2$ at 1.9 K obtained from the spectra by adopting the formula (1). In LF, the muon-spins rotate around the total field of the internal field and LF at the muon site. The spectra changes with increasing $H_{\rm LF}$, which is well represented by the increase of the $A_2$ fraction in the formula (1). This means that the muon-spin is gradually decoupled by LF from the static internal field which causes the fast depolarization below $T_{\rm MI}$. The static internal field $H_{\rm int}$ at this temperature was estimated from the $H_{\rm LF}$ dependence of $A_2$ by using the formula

$$A_2 = \frac{3}{4} - \frac{1}{4x^2} + \frac{(x^2-1)^2}{16x^3} \ln \frac{(x+1)^2}{(x-1)^2}, \qquad (2)$$

where $x = H_{\rm LF}/H_{\rm int}$. This formula is derived from the assumption that $H_{\rm int}$ has a unique magnitude but have random directions to $H_{\rm LF}$ [28, 29]. Then, $H_{\rm int} \simeq 65$ Oe was evaluated at this temperature. This field is much higher than that expected for nuclear dipole moments.

Figure 5 shows $T$ dependence of $H_{\rm int}$ obtained by the same procedures. The $H_{\rm int}$ appears below $T_{\rm MI}$. Together with the ZF-$\mu$SR results which suggest the appearance of a magnetic transition at $T_{\rm MI}$, these results indicate that the static internal field appears below $T_{\rm MI}$ but not below $T_{\rm N}$. This concludes that the ordering below $T_{\rm MI}$ is not a non-magnetic AFQ ordering but a magnetic one with the TRS breakdown.

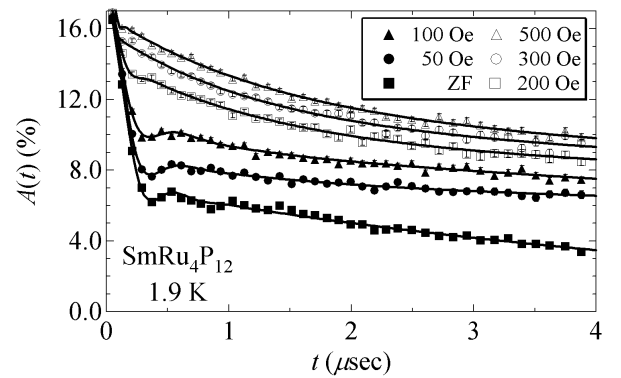

FIG. 3: LF-$\mu$SR time spectra at 1.9 K in several longitudinal fields. The solid lines show the best-fit results by the formula (1).

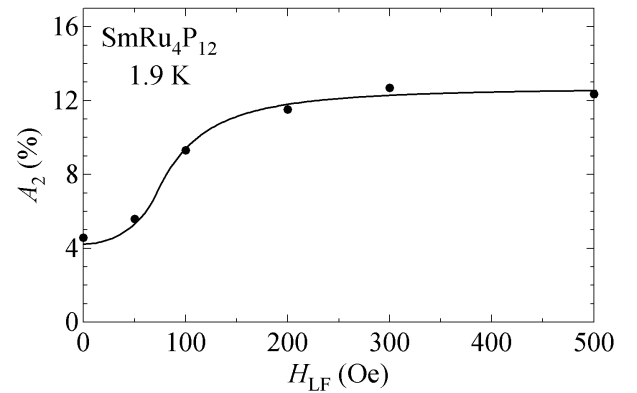

FIG. 4: Longitudinal field dependence of $A_2$ at 1.9 K obtained from the best-fit results by the formula (1) shown as the solid lines in FIG. 3. The solid line shows the best-fit result by the formula (2).

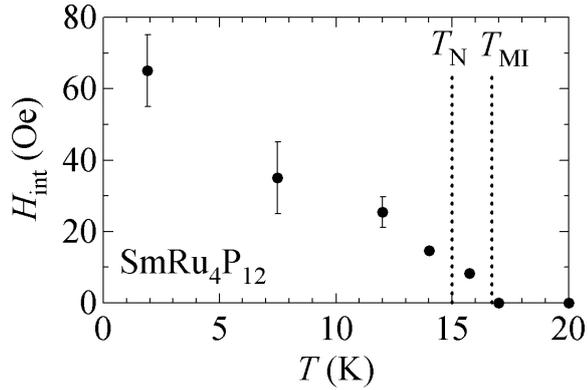

FIG. 5: Temperature dependence of $H_{int}$ obtained from the decouplings of the LF-$\mu$SR spectra.

The TRS breakdown in $NpO_2$ was also confirmed by observing the muon-spin precession below the transition temperature of 25 K [30]. Compared with the case of $NpO_2$, the precession in $SmRu_4P_{12}$ shows a fast dumping, which means that the spin alignment is not so coherent. Although the origin of the low coherency is not clear at the moment, it could be guessed that the fluctuating Sm moments smear out the precession.

A clear anomaly was not observed around $T_N$, which is consistent with other experiments [10, 12, 17, 19, 20]. From a group-theoretical consideration based on the Landau theory, this ambiguous transition can be explained by the coupling of an octupolar ($\Gamma_{5u}$) and a dipolar ($\Gamma_{4u}$) order parameter [20]. The strength of the coupling depends on the separation of the transition temperatures. This is supported by a result that $\Gamma_{5u}$ and $\Gamma_{4u}$ belong to the same irreducible representation in the $T_h$ symmetry. In ZF and low fields, the coupling can smear out the anomaly at $T_N$ because the separation between each transition temperature is small [20]. In this case, the anomaly around $T_N$ becomes ambiguous. On the other hand, with increasing field, the anomaly around $T_N$ becomes clear. Hence, the present result is consistent with the expected behavior in the octupolar-dipolar (AFM) scenario [20]. In addition, it is noted that a magnetic dipole moment has been excluded as a candidate for the order parameter below $T_{MI}$ because a discontinuous transition is expected at $T_N$ in this case [20]. Therefore, the probable order parameter for this M-I transition is a magnetic octupolar ordering.

It is also pointed out from multiorbital Anderson model calculations that a $\Gamma_{5u}$ and $\Gamma_{4u}$ octupolar fluctuation can become significant in Sm-based filled skutterudite systems [31]. In the case of $NpO_2$, it is reported that the hyperfine interactions at O sites obtained from $^{17}$O-NMR can be well explained by a longitudinal triple-$\vec{q}$ multipolar structure [22, 23]. In order to establish the octupolar scenario in $SmRu_4P_{12}$, angle resolved NMR by using single crystal and neutron diffraction experiments by using a sample without an isotope of $^{149}$Sm nucleus are strongly required in future works. In conjunction with these results, an octupolar structure can be specified in $SmRu_4P_{12}$.

## IV. SUMMARY

In summary, the ZF- and LF-$\mu$SR measurements on $SmRu_4P_{12}$ have been performed. The $T$ dependence of both $A_2$ and $\lambda_2$ obtained from the ZF-$\mu$SR spectra have the anomaly around $T_{MI}$. The appearance of the static internal field below $T_{MI}$ was confirmed from the decouplings of the LF-$\mu$SR spectra. These results are direct evidences for the magnetic ordering with the TRS breakdown below $T_{MI}$ but not a non-magnetic AFQ ordering, which supports a scenario that a magnetic octupolar ordering occurs below $T_{MI}$ in $SmRu_4P_{12}$.

### Acknowledgments

The authors would like to thank M. Yoshizawa, K. Matsuhira, T. Hotta and H. Harima for their useful discussions. This work was supported by the Sasakawa Scientific Research Grant from The Japan Science Society, the Toray Science and Technology, the Joint Research Project and Grant-in-Aid for the Japan Society for the Promotion of Science, and Grant-in-Aid for Scientific Research in Priority Area "Skutterudite" (No. 15072201) of the Ministry of Education, Culture, Sports, Science and Technology, Japan.